# Uncertain research country rankings. Should we continue producing uncertain rankings?

Alonso Rodríguez-Navarro

*Departamento de Biotecnología-Biología Vegetal, Universidad Politécnica de Madrid, Avenida Puerta de Hierro 2, 28040, Madrid, Spain*

*E-mail address: alonso.rodriguez@upm.es.*

ORCID 0000-0003-0462-223X

**Abstract**

**Purpose:** Citation-based assessments of countries' research capabilities often misrepresent their ability to achieve breakthrough advancements. These assessments commonly classify Japan as a developing country, which contradicts its prominent scientific standing. The purpose of this study is to investigate the underlying causes of such inaccurate assessments and to propose methods for conducting more reliable evaluations.

**Design/methodology/approach**: The study evaluates the effectiveness of top-percentile citation metrics as indicators of breakthrough research. Using case studies of selected countries and research topics, the study examines how deviations from lognormal citation distributions impact the accuracy of these percentile indicators. A similar analysis is conducted using university data from the Leiden Ranking to investigate citation distribution deviations at the institutional level.

**Findings**: The study finds that inflated lower tails in citation distributions lead to undervaluation of research capabilities in advanced technological countries, as captured by some percentile indicators. Conversely, research-intensive universities exhibit the opposite trend: a reduced lower tail relative to the upper tail, which causes percentile indicators to overestimate their actual research capacity.

**Research limitations**: The descriptions are mathematical facts that are self-evident.

**Practical implications**: Due to variations in citation patterns across countries and institutions, the $P_{top\ 10\%}/P$ and $P_{top\ 1\%}/P$ ratios are not universal predictors of breakthrough research. Evaluations should move away from these metrics. Relying on inappropriate citation-based measures could lead to poor decision-making in research policy, undermining the effectiveness of research strategies and their outcomes.

## 1. Introduction

Since the mid-20th century, science and technology have emerged as key drivers of economic growth (Godin, 2004). As a result, research has evolved from a tool for advancing knowledge to a vital factor in sustaining national wealth. Consequently, research assessment has become an essential instrument for managing systems of critical importance within nations. Peter Drucker, widely regarded as the father of modern management (Drucker, 1954), is credited with the statement: "You can't manage what you can't measure." This opinion aligns with the more recent observation that "Government policymakers, corporate research managers, and university administrators need valid and reliable S&T indicators" (Garfield & Welljams-Dorof, 1992, p. 321).

In response to political demands (Godin, 2006) and academic initiatives (Garfield, 1955; Merton, 1979; Narin, 1976; Price, 1965), a growing number of countries, institutions, and researchers have turned to bibliometric measures to assess the scientific and technological achievements of countries and organizations (Aksnes et al., 2019; Godin, 2003; Mingers & Leydesdorff, 2015; Waltman, 2016). Leading institutions, such as the European Commission (EC) (European Commission, 2018, 2020, 2022, 2024) or the US National Science Board (NSB) (National Science Board, 2018, 2020, 2022, 2024), regularly publish reports that assess the research performance of various countries.

Despite significant efforts by policymakers and academics to develop reliable research indicators, there is growing evidence that many commonly used indicators can be misleading. A notable example is Japan, which, despite its robust research output, is often ranked as a developing country in research assessments (Pendlebury, 2020). For instance, the latest EC report, using an indicator deemed representative of scientific excellence, places Japan at the level of underdeveloped countries (European Commission, 2024, Figure 3.1-22, p. 175); Similarly, the NSB, while adopting more cautious language, has consistently ranked Japan and India similarly, behind Germany, the UK, and the USA from 2003 to 2020 (National Science Board, 2024, Figure 19, p. 23). While misclassifications may go unnoticed in other countries, Japan's case is particularly evident. Over the 21st century, 17 Japanese researchers have been awarded Nobel Prizes, and a more detailed study places Japan at the level of the UK and Germany in terms of Nobel laureates (Schlagberger et al., 2016). Furthermore, Japan surpasses these countries in terms of triadic patents (Rodríguez-Navarro & Brito, 2022). Thus, evaluation methods that categorize Japan as a scientifically underdeveloped country are fundamentally unreliable.

### 1.1. Citation metrics for research assessments

Bibliometric assessments based on the number of publications and citation (Waltman, 2016; Wildgaard et al., 2014) are the simplest and most commonly used method to evaluate the scientific success of countries and institutions, although these methods are not suitable for assessing individual papers or small sets of publications (Aksnes et al., 2019). Citation metrics can be divided into three categories: citation-only metrics, citation-based rank metrics, and combined metrics. When applying citation metrics across different disciplines and publication years, it is essential to normalize them to account for variations in citation practices between fields. This principle has long been recognized: comparing two publications from different research fields requires normalizing their citation counts (Mcalister et al., 1983; Schubert & Braun, 1996).

Citation-only metrics rely on the total number of citations to assess the impact of papers. These metrics are used to evaluate the performance of countries and institutions by considering the citation counts of all their publications. This number of citations can also be used in combination with the number of papers. The most basic of these metrics is the mean number of citations per publication, similar to the journal impact factor (Garfield, 1999). However, more sophisticated formulas have also been proposed (Docampo & Besoule, 2019; Prathap, 2010; Waltman et al., 2011).

Citation-based rank metrics are non-parametric indicators (Conover & Iman, 1981) derived from citation counts. In this approach, papers from specific years, research disciplines, or topics are ranked according to the number of citations they have received, starting with the most cited. A similar ranking is applied to the publications of institutions or countries. As a result, each publication receives two ranks—one in the global list and one in the local list—and these ranks are used for evaluation. A widely used metrics in this category are the top percentile metrics, where the global list is segmented into top percentiles (e.g., top 10% or 1%) (Bornmann et al., 2013; Waltman

& Schreiber, 2013). Then, the percentile indicator is the number of papers in the local list that appears in the corresponding global percentile, which coincides with the local rank of the least cited paper in the segmented global list—papers with the same number of citations raise an issue that has technical solutions (Waltman & Schreiber, 2013). This method is intuitive because the concept of success is associated with a high number of citations and, consequently, with a high number of papers in a narrow top percentile. Other metrics of this type are based on the two ranks, either using most of papers in the list, as in the $e_p$ index (Rodríguez-Navarro & Brito, 2018b), or only a few papers in the top positions in the lists, as in the *Rk* index (Rodríguez-Navarro & Brito, 2024b).

The third type of metric combines both ranks and citation counts. A widely known indicator of this type is the *h*-index, which has generated an abundant bibliography about it and its derivatives (Bihari et al., 2021; Bornmann & Daniel, 2007). The *h*-index is the simplest combination of parametric and non-parametric statistics; it is determined when, in the list of papers, the rank and the number of citations of are equal (Hirsch, 2005). However, it is not an accurate indicator (Brito & Rodríguez-Navarro, 2021; Waltman & van Eck, 2012).

### 1.2. Research indicators and the ideal citation model

The primary aim of research evaluations is often to assess a country's or institution's contribution to advancing knowledge or its ability to produce scientific knowledge that can be translated into innovations within the productive system. However, the concrete realization of these concepts remains somewhat ambiguous. It is likely that all publications in internationally recognized scientific journals contribute in some way to the progress of knowledge, and in technological fields, many papers may eventually lead to innovations, or at least contribute to them.

When focusing on evaluating disruptive research, the problem becomes more manageable by recognizing that scientific progress is not purely incremental. Instead, it alternates between large quantities of incremental (normal) science and much smaller amounts of revolutionary science or breakthrough discoveries (Kuhn, 1970). Under this model, research evaluations can be particularly focused on publications reporting breakthrough discoveries, which are often identified by their high citation counts (Bornmann et al., 2013; Min et al., 2021; Schneider & Costas, 2017; Wuestman et al., 2020). However, it is important to note that not all breakthrough papers are highly cited (Hu & Rousseau, 2019), and not all highly cited papers necessarily represent scientific breakthroughs (Garfield, 1973).

To define the appropriate citation-based metrics for breakthrough evaluations, it is essential to establish a citation threshold, and this threshold presents a challenge since it typically falls within the top 0.01% or 0.02% citation percentile (Bornmann et al., 2018; Poege et al., 2019). Evaluating research at such a stringent level is challenging, particularly for smaller countries, research institutions, or even large countries with less competitive research systems, which may take years to produce a paper that reaches this citation level. To address this issue, research evaluations have often been based on the number of publications in the top 10% or 1% citation percentile in a given field or topic (Waltman, 2013 #1587).

Given this context, the emerging question is whether evaluations based on the top 10% or 1% citation percentile can predict the occurrence of papers in the top 0.01% or 0.02% citation percentile. To answer this, we must distinguish between cases where citations follow or deviate from the ideal model. According to this model, the rank distribution of a country's or institution's publications (local versus global rank plot; Section 1.1) follows a power law (Rodríguez-Navarro & Brito, 2018a). This relationship allows for the calculation of the number of publications in any top percentile based solely on the total number of papers and the number of publications in a single top percentile (Rodríguez-Navarro & Brito, 2021b), or alternatively, calculating the probability of having a publication in a narrow top percentile (Rodríguez-Navarro & Brito, 2019). However, just as the ideal gas law does not perfectly describe all gas behaviors, many countries and institutions deviate from this ideal model. Such deviations occur when the citation distribution of a country's or institution's publications differs significantly from the global citation distribution (Rodríguez-Navarro, 2024).

In research aimed at pushing the boundaries of knowledge, the lognormal citation distribution (Radicchi et al., 2008) is well-supported (Golosovsky, 2021) and has the advantage of a relatively constant $\sigma$ parameter, typically around 1.0–1.2 (Radicchi et al., 2008; Rodríguez-Navarro & Brito, 2018a; Thelwall, 2016; Viiu, 2018).

However, research objectives can vary, and deviations from the lognormal distribution are common (Perianes-Rodriguez & Ruiz-Castillo, 2016; Rodríguez-Navarro & Brito, 2024a; Shahmandi et al., 2020; Waltman et al., 2012), often due to a large number of uncited publications (Rodríguez-Navarro & Brito, 2024a; Waltman et al., 2012). Considering the diversity of possible cases, a previous study suggests that defining a research system requires three indicators in addition to the total number of publications: for the extreme of the upper tail, for lowly cited papers, and for the efficiency of the system in the top 10% cited papers, approximately (Rodríguez-Navarro, 2024).

### 1.3. Uncertain research rankings and evaluations

Most indicators used in research assessments are statistical data or are calculated from statistical data. In certain cases, their role as indicators of research performance is assumed, while in others, it is validated through peer reviews in university assessments (Rodríguez-Navarro & Brito, 2020; Thelwall et al., 2023; Traag & Waltman, 2019). Laying aside the accuracy of peer-review assessments (Abramo et al., 2013), the stringency of these evaluations cannot be high, at the level of the top 10% or 1% of cited papers (Rodríguez-Navarro & Brito, 2020). This represents the highest level of stringency attainable through by peer review, which requires that the number of reviewed papers is sufficiently high. Peer review cannot effectively assess the likelihood of a department or university producing breakthrough papers that rank within the top 0.01%, this level of granularity can only be derived through bibliometric analysis.

Assuming the notion that reliable descriptions of research systems require three indicators in addition to the number of papers (Rodríguez-Navarro, 2024), the arising question is: What kind of information do research assessments based on a single indicator provide? Focusing on the number of top 10% or 1% cited papers (henceforth, I will use the notation $P_{top\ x\%}$ for the number of paper in top x% percentile and P for the total number of papers), as well as on breakthrough contributions, the conclusion is worrying. Using $P_{top\ 0.01\%}$ as a reference, only one paper in every 1000 within $P_{top\ 10\%}$ and one in every 100 within $P_{top\ 1\%}$ reaches this level of impact. If citation fits the ideal model, the contribution at the breakthrough level ($P_{top\ 0.01\%}$) can be computed using the ideal power law (Rodríguez-Navarro & Brito, 2019), but this is not applicable in other cases. When considering research efficiency based on size-independent indicators, if the research system does not adhere to the ideal model, parameters such as $P_{top\ 10\%}/P$ and $P_{top\ 1\%}/P$ can be highly misleading, as demonstrated in the case of Japan above described (Section 1).

Testing whether citation patterns conform the ideal model is straightforward because, in this case, the $P_{top\ 10\%}/P$, $P_{top\ 5\%}/ P_{top\ 50\%}$, and $P_{top\ 1\%}/P_{top\ 10\%}$ ratios are equal (Rodríguez-Navarro & Brito, 2019). Using this test, in the research topic of graphene, only 50% of the countries adhere to the ideal model (Rodríguez-Navarro, 2024). This implies that, in this research topic, 50% of country's research evaluations based on $P_{top\ 10\%}/P$ or $P_{top\ 1\%}/P$ are misleading.

This raises the issue that in common research rankings or evaluations, selecting a specific country provides no indication of whether the evaluation is accurate, approximate, or misleading. A simple calculation could resolve this uncertainty, but, to the best of my knowledge, such calculations are not currently included in research country rankings or evaluations.

## 2. Aim of this study

Given these concerns, this study aims to complement a previous investigation into deviations from the lognormal citation distribution (Rodríguez-Navarro, 2024). The first part of the study examines countries across three research topics—solar cells, semiconductors, and immunity—by analyzing a limited sample of countries. These case studies highlight the diversity of deviations in both the upper and lower extremes of the citation distribution and their effect on the ideal rank power law. The second part leverages data from the Leiden Ranking to assess the alignment or deviation of university research from the ideal model.

## 3. Materials and Methods

To investigate the deviations described above in countries' publications, global and country papers in the selected topic were ordered based on their number of citations, starting with the one with the highest number of citations. The global ranks of country papers were obtained after their identification in the global list. Scatterplots of ranks are presented in double logarithmic scales where a power law is transformed into a straight line. Deviations from the ideal power law are easily detected as deviations from a straight line. For deviations in the lower tail, I constructed histograms using logarithmic binning, including the numbers of papers with zero, one, and two citations. In some figures, the histogram of a synthetic series of lognormally distributed random numbers is included to guide the eye. These series were generated as described previously (Rodríguez-Navarro & Brito, 2018a).

To investigate the deviation from the ideal power law in universities, I used the data reported in the CWTS Leiden Ranking (Olensky et al., 2015; Waltman, Calero-Medina, et al., 2012) for the year 2023 (https://zenodo.org/records/8027120). In this case, if the global ranks of the university publications follow the ideal power law, the top percentiles reported in the Leiden Ranking meet the equalities $P_{top\ 10\%}/P = P_{top\ 5\%}/P_{top\ 50\%} = P_{top\ 1\%}/P_{top\ 10\%}$ (Rodríguez-Navarro & Brito, 2019). The reported data are the means of four periods of evaluation, from 2015–2018 to 2018–2021, using fractional counting. Universities with less than 10 papers in the top 1% most cited papers ($P_{top\ 1\%}$) in any of the four evaluation periods were eliminated.

Citations in the topics of solar cells/photovoltaics, semiconductors, and immunity were obtained from Clarivate Web of Sciences. Searches were performed making use of the Advanced Search tool in the Web of Science Core Collection, edition: Science Citation Index Expanded. The searches were limited to articles (DT=) using a four-year publication window (2014–2017) and counting the citations in a four-year window (2019–2022). The global search was limited to the 75 most productive countries (Clarivate InCites). Domestic searches included one country and excluded the other 74. Focusing on domestic papers is advantageous in this study because, in general terms, deviations in the lower tail are more evident in domestic than in international collaborative papers (unpublished results).

## 4. Results

### 4.1. The upper tail of citation distributions

The measurement or estimation of the contribution or capacity of countries and institutions in pushing the boundaries of knowledge should be one of the most important aims of research assessment. In every topic or research field evaluated, this information can be obtained by studying the very highly cited papers that constitute the

**Table 1**. Basic description of scientific publications in terms of P, $P_{top\ 10\%}/P$, and $P_{top\ 1\%}/P_{top\ 10\%}$ in 18 selected cases

| | Solar cells | | | Semiconductors | | | Immunity | | |
|---|---|---|---|---|---|---|---|---|---|
| **Country** | **P** | **$P_{top\ 10\%}$/ P** | **$P_{top\ 1\%}$/ $P_{top\ 10\%}$** | **P** | **$P_{top\ 10\%}$/ P** | **$P_{top\ 1\%}$/ $P_{top\ 10\%}$** | **P** | **$P_{top\ 10\%}$/ P** | **$P_{top\ 1\%}$/ $P_{top\ 10\%}$** |
| Global | 61202 | | | 58081 | | | 42586 | | |
| Germany | 1834 | 0.077 | 0.063 | 1748 | 0.065 | 0.061 | 1177 | 0.091 | 0.056 |
| India | 3138 | 0.052 | N/C | 3056 | 0.046 | 0.056 | 962 | 0.021 | N/C |
| Japan | 2742 | 0.059 | 0.092 | 3167 | 0.058 | 0.087 | 1491 | 0.057 | 0.106 |
| South Korea | 4193 | 0.068 | 0.102 | 2668 | 0.066 | 0.068 | 1088 | 0.053 | N/C |
| UK | 892 | 0.182 | 0.204 | 810 | 0.115 | 0.204 | 893 | 0.103 | 0.076 |
| USA | 5474 | 0.146 | 0.141 | 6004 | 0.149 | 0.111 | 7946 | 0.145 | 0.122 |

Publication window: 2014–2017; citation window: 2019–2022. N/C, not calculated because $P_{top\ 1\%}$ is too low extreme upper tail of the citation distribution of global papers. To address the question of whether this extreme of the upper tail conforms to or deviates from the ideal model, double rank analysis is the most convenient approach.

To perform this analysis, I chose three important research topics: solar cells/photovoltaics, semiconductors, and immunity, which have experienced rapid evolution in recent years. In two of these topics, solar cells and semiconductors, incremental innovations have proven extremely important, whereas this type of innovation holds less significance in immunity. I then selected six countries based on their distinct research system models. Table 1 provides a basic description of these 18 cases in terms of P, $P_{top\ 10\%}/P$, and $P_{top\ 1\%}/P_{top\ 10\%}$. In the absence of deviations from the ideal rank power law, the ratios $P_{top\ 10\%}/P$ and $P_{top\ 1\%}/P_{top\ 10\%}$ should be equal (Rodríguez-Navarro & Brito, 2019). This is evident in a few cases, such as the USA in solar cells, and Germany and South Korea in semiconductors. In other cases, differences are high in either direction, such as Germany and Japan in immunity. Overall, describing a general pattern for the comparisons of these ratios proves challenging: each country and topic may be different. The existence of high deviations from equality in these ratios is noteworthy, as it indicates that the use of single indicators like $P_{top\ 10\%}/P$ to assess contributions to pushing the boundaries of knowledge can be misleading.

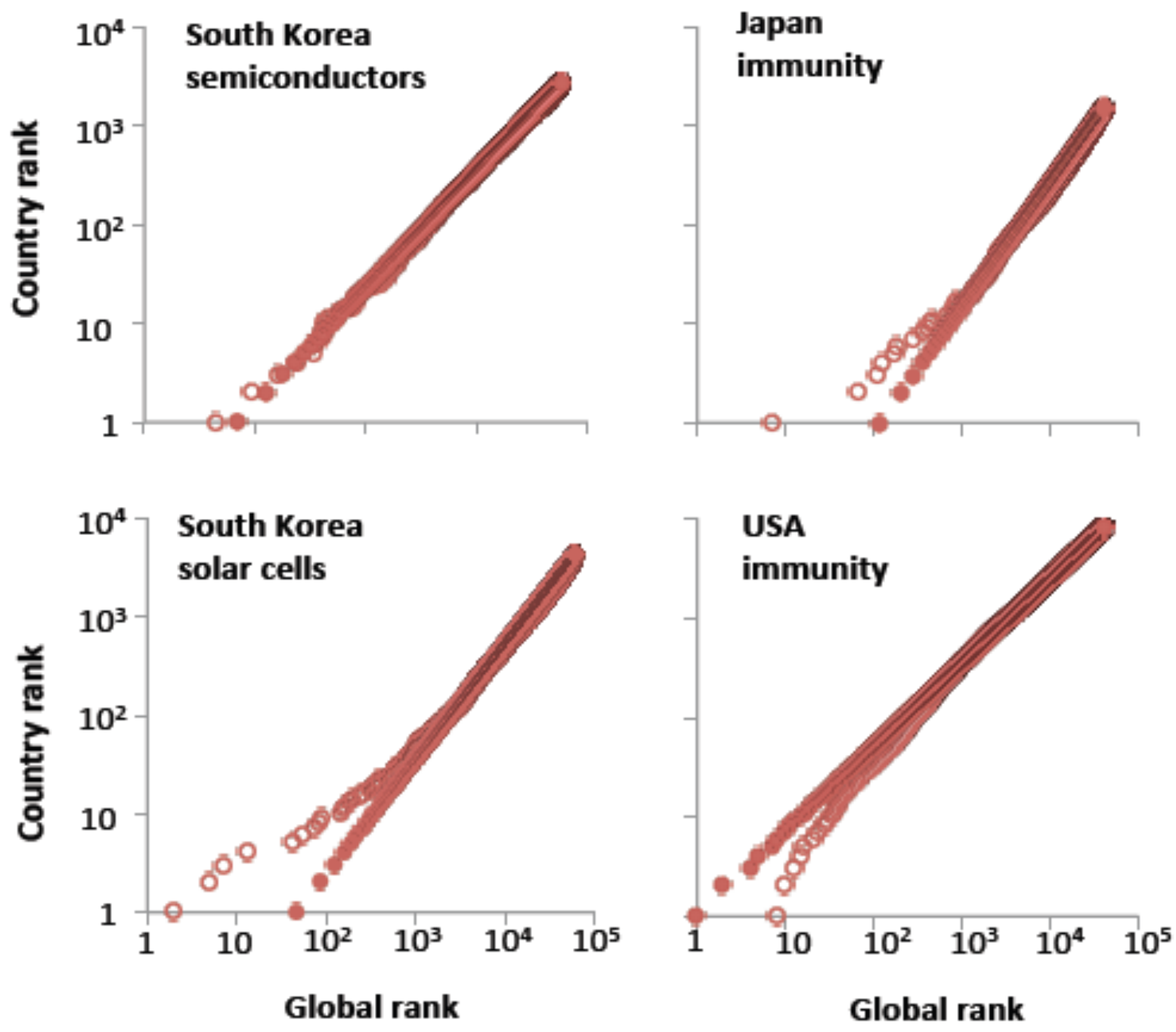

Figure 1. Country versus global ranks of domestic papers. Comparison of actual global ranks (open circles) with those calculated for the power law defined by P and $P_{top\ 10\%}$ (closed circles). Publication window: 2014–2017; citation window: 2019–2022

For double rank analysis, the country rank (ordinate axis) is plotted versus the global rank (abscissa axis; Rodríguez-Navarro & Brito, 2018a). Then, to study deviations, the double rank plot of the actual data is compared to the plot of data mathematically generated. Using as a model the power law defined by P and $P_{top\ 10\%}$, the comparison reveals three different patterns (Figure 1): (i) the reference and actual plots practically overlap (South Korea in semiconductors; also Table 1), ensuring the accuracy of the evaluations based on P and $P_{top\ 10\%}$; (ii) the most cited papers have smaller global ranks than expected from P and $P_{top\ 10\%}$ (Japan in immunity and South Korea in solar cells; also Table 1), implying that evaluations based on P and $P_{top\ 10\%}$ undervalue the research capacity; (iii) the most cited papers have larger global ranks than expected from P and $P_{top\ 10\%}$ (USA in immunity; also Table 1), implying that evaluations based on P and $P_{top\ 10\%}$ overvalue the research capacity. These last two patterns imply that, in these cases, P and $P_{top\ 10\%}$ should not be used to predict the contribution to pushing the boundaries of knowledge.

The plots in Figure 1 demonstrate that, considering the total number of papers (P), the extreme upper tail cannot be predicted. This raises the question of whether it could be predicted using only the top 10% cited papers. Figure 2 shows that even with this restriction, the extreme upper tail is not predicted in some cases. This figure uses the power law defined by $P_{top\ 10\%}$ and $P_{top\ 1\%}$ instead of P and $P_{top\ 10\%}$, as in Figure 1, in two cases: South Korea in solar cells and the USA in immunity. In both cases, the deviations between the actual and calculated data persist, although they are reduced.

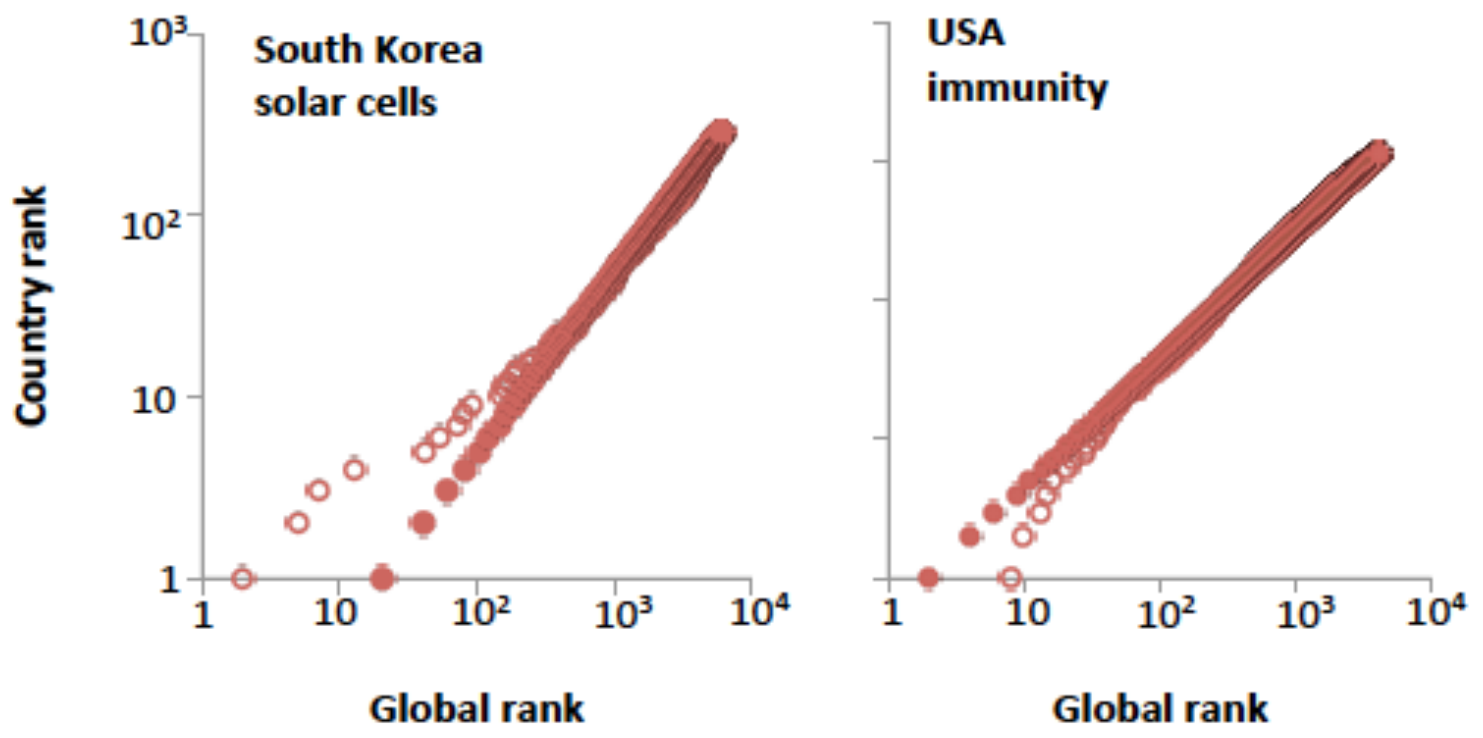

**Figure 2.** Country versus global ranks of domestic papers. Comparison of actual global ranks (open circles) with those calculated for the power law defined by $P_{top\ 10\%}$ and $P_{top\ 1\%}$ (closed circles). Publication and citation windows, as in Figure 1

A simple conclusion can be drawn from the study of the extreme upper tail of the citation distribution. It implies that, as a general rule, the contribution of countries to pushing the boundaries of knowledge cannot be predicted from citation metrics based on papers that are not in the extreme upper tail. Restricting the rank analysis to the top 10% of most cited papers improves the prediction a little, but not sufficiently.

### 4.2. Inflated lower tail in citation distributions

Another difficulty in describing the research systems of countries arises from deviations from the lognormal distribution in the less-cited papers (Rodríguez-Navarro, 2024; Rodríguez-Navarro & Brito, 2024a). To study the lower tail, the double rank approach has little sensitivity, and I used a logarithmic binning approach, including the number of papers with 0, 1, and 2 citations.

A first analysis of the global papers in the three topics studied here (Figure 3) shows that their citation distributions deviate notably from a lognormal distribution, which fits reasonably well the middle of the distribution and its upper tail. However, the three topics are different. In immunity, the distribution can be described as lognormal with a minor excess of papers that are uncited or have one citation. In solar cells and semiconductors, the distributions are more complex, showing an excess of papers in all bins to the left of the mode in the histogram.

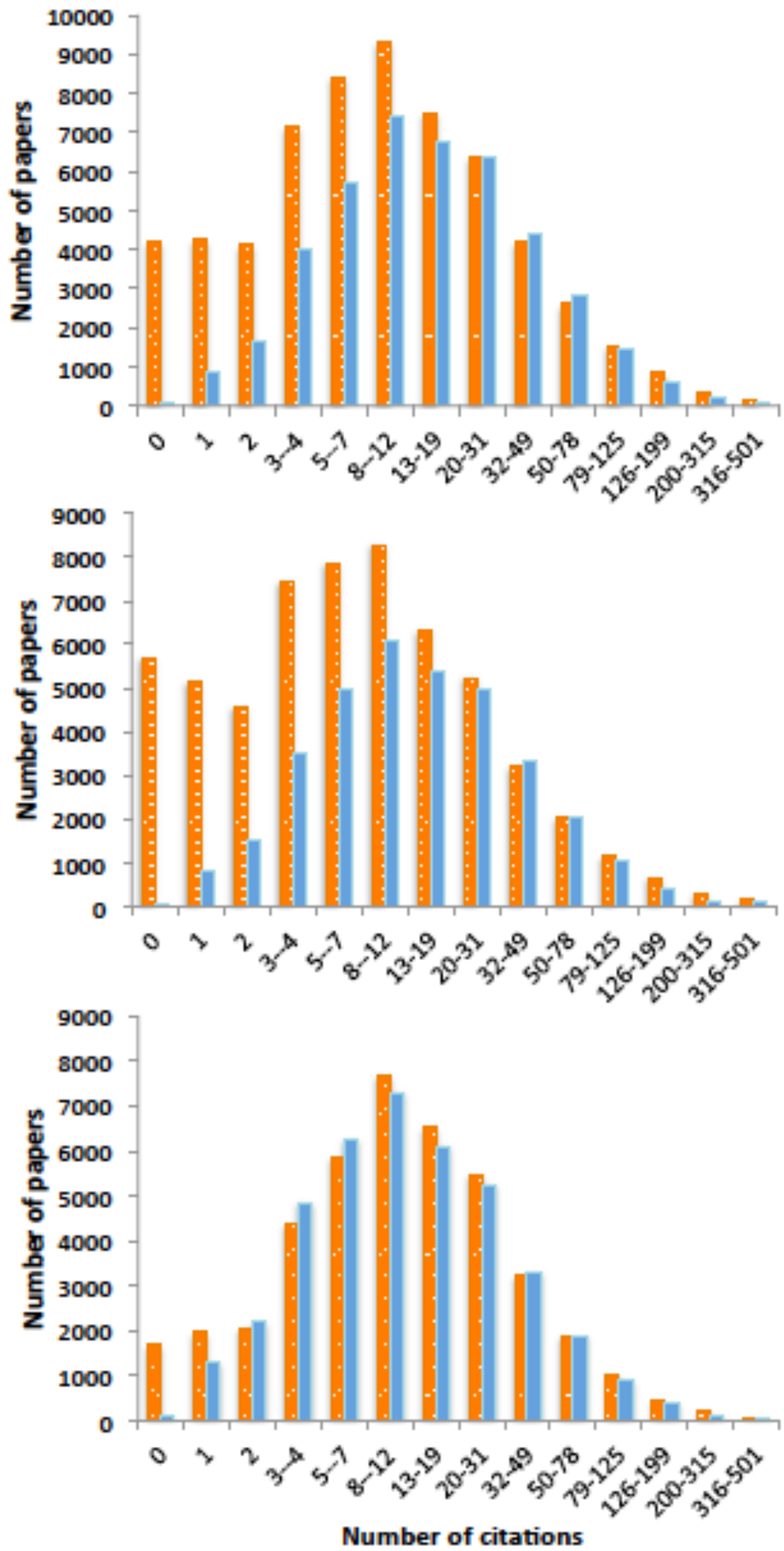

**Figure 3**. Histogram with logarithmic binning of citations to world publications in the scientific topics of solar cells, semiconductors, and immunity (from top to bottom). Blue bars correspond to a lognormal synthetic series to guide the eye. Publication and citation windows, as in Figure 1

Next, I investigated the countries' variability in the lower tail. Figure 4 shows the logarithmic binning histograms of three countries—UK, Japan, and South Korea—

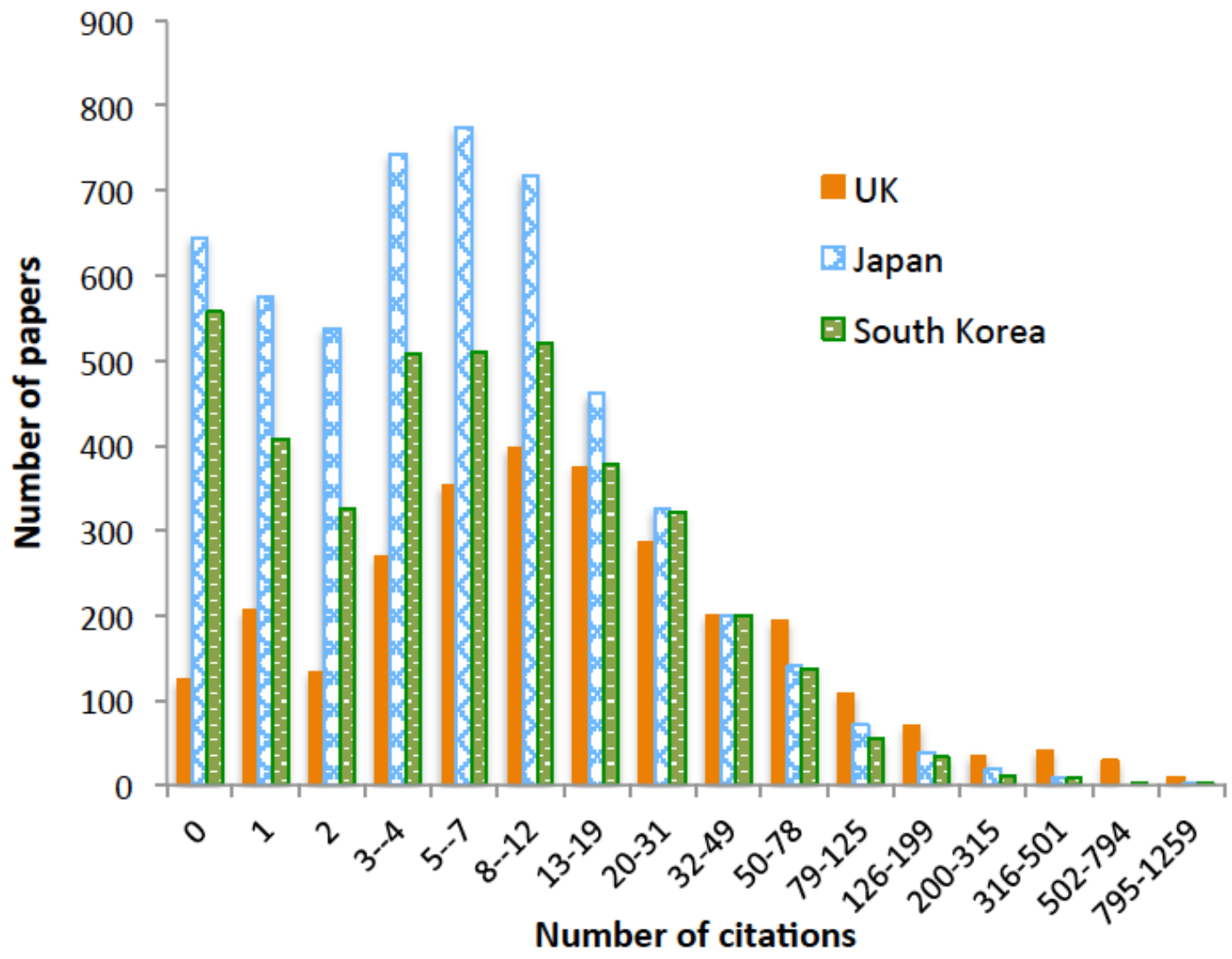

**Figure 4.** Histogram with logarithmic binning of citations to publications from the UK, Japan, and South Korea on the scientific topic of solar cells. Publication and citation windows, as in Figure 1

in the topic of solar cells. For easier comparison, the distributions of the UK and Japan have been scaled up to the level of South Korea in the 32–49 bin. While the UK distribution has a moderate excess of papers with zero and one citation, in Japan and South Korea, the deviations are significant and affect a large proportion of the papers. Notably, above the 13–19 bin, the three distributions follow similar patterns, except for the extreme upper tail, which is in contrast with the large deviations in the lower tail of the distribution.

Considering the differences between topics presented in Figure 3 and the discrepancies between countries shown in Figure 4, it appears that the likelihood of identifying a universal research indicator that effectively captures the research differences between the UK, Japan, and South Korea is extremely low. In Figure 4, the substantial disparity in the number of papers with zero or one citation between the UK (which has low counts) and Japan and South Korea (which have high counts) suggests a pronounced differential effect on size-independent indicators derived by dividing by P. For instance, the $P_{top\ 10\%}/P$ ratios for two countries with identical upper tails could differ significantly. The assumption that papers with zero or one citation hold no research value, which is implicit in the $P_{top\ 10\%}/P$ metric for predicting research performance, is a perilous one.

### 4.3. Uncertainties in universities

To further investigate the deviations of citation distributions from the ideal rank power law, I utilized the extensive data provided by the Leiden Ranking. As already mentioned, if the double-rank power law model applies, the $P_{top\ 10\%}/P$, $P_{top\ 5\%}/P_{top\ 50\%}$, and $P_{top\ 1\%}/P_{top\ 10\%}$ ratios should be equal (Rodríguez-Navarro & Brito, 2019).

To test this requirement, I selected two fields, *Physical sciences and engineering* and *Biomedical and health sciences*, the data from four evaluation periods, and universities in which $P_{top\ 1\%}$ is at least 10 using fractional counting. The selection meeting the last criterion included 311 universities in 30 countries in *Physical sciences and engineering*, and 310 universities in 28 countries in *Biomedical and health sciences* (Supplementary Tables S1 and S2). The scatter plots of $P_{top\ 5\%}/P_{top\ 50\%}$ and $P_{top\ 1\%}/P_{top\ 10\%}$ versus $P_{top\ 10\%}/P$ show good correlations in the case of $P_{top\ 5\%}/P_{top\ 50\%}$ but worse in the case of $P_{top\ 1\%}/P_{top\ 10\%}$. In the field of *Physical sciences and engineering* the dispersion of the data is higher than in the field of *Biomedical and health sciences* (Figure 5).

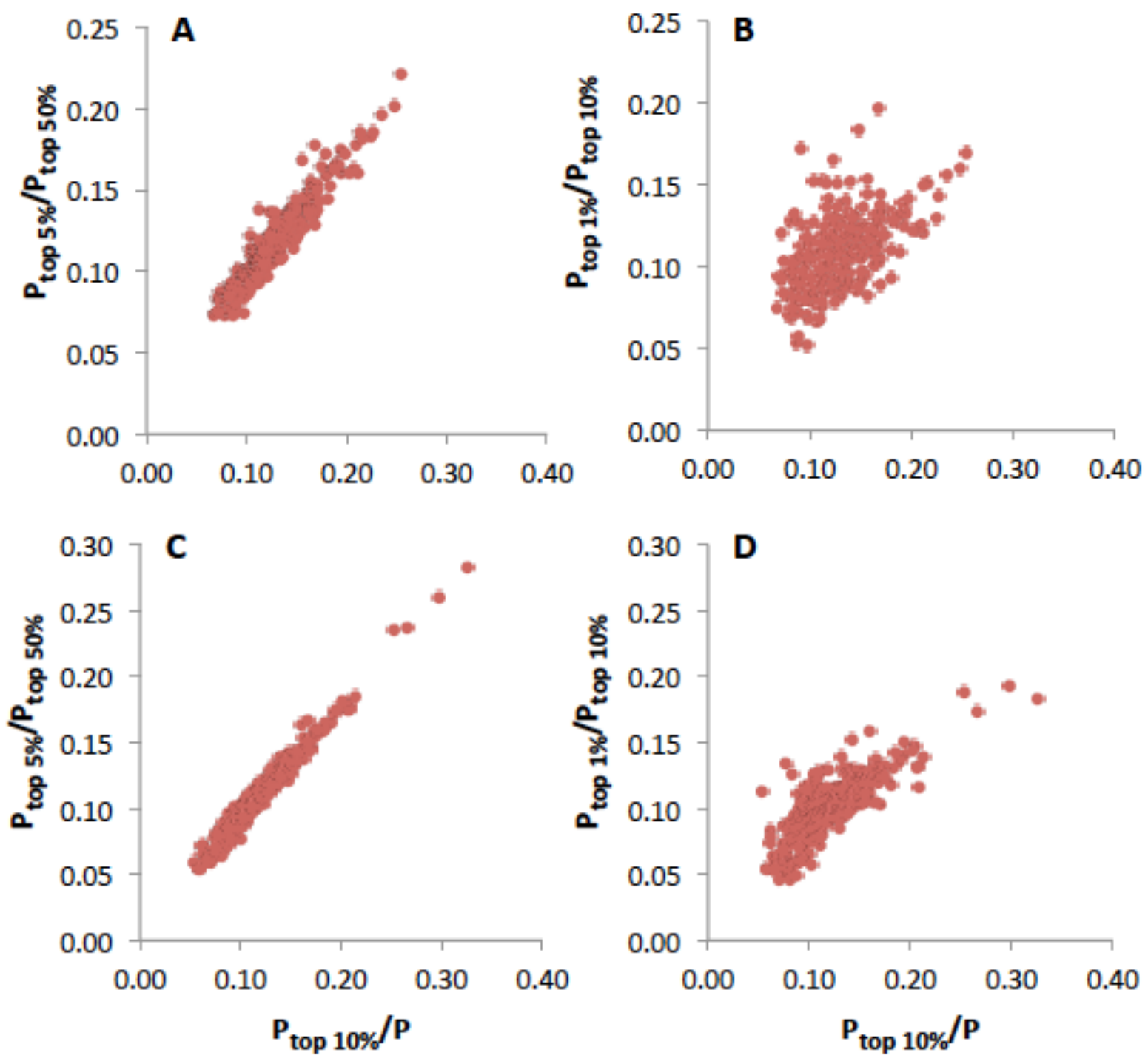

Figure 5. Plots of $P_{top\ 5\%}/P_{top\ 5\%}$ and $P_{top\ 1\%}/P_{top\ 10\%}$ versus $P_{top\ 10\%}/P$ in selected universities. Data from the CWTS Leiden Ranking 2023, means of four evaluation periods, and fractional counting. A and B panels: Physical and engineering sciences; C and D panels: Biomedical and health sciences. Evaluation periods and selection of universities are described in the text

**Table 2**. Classification of universities attending to the variation of the $P_{top\ 10\%}/P$, $P_{top\ 5\%}/P_{top\ 50\%}$, and $P_{top\ 1\%}/P_{top\ 10\%}$ ratios: stable (Type A), decreasing (Type B), or increasing (Type C). Field of Physical sciences and engineering

| University | Country | $P_{top\ 10\%}/P$ | $P_{top\ 5\%}/P_{top\ 50\%}$ | $P_{top\ 1\%}/P_{top\ 10\%}$ |
|---|---|---|---|---|
| **Type A** | | | | |
| Pennsylvania State University | United States | 0.138 | 0.129 | 0.117 |
| Florida State University | United States | 0.124 | 0.121 | 0.130 |
| University of Virginia | United States | 0.121 | 0.118 | 0.118 |
| University of Nottingham | United Kingdom | 0.141 | 0.125 | 0.120 |
| University of Exeter | United Kingdom | 0.127 | 0.108 | 0.125 |
| Chongqing University | China | 0.113 | 0.106 | 0.109 |
| Hohai University | China | 0.103 | 0.113 | 0.089 |
| University of Queensland | Australia | 0.140 | 0.131 | 0.151 |
| University of Bordeaux | France | 0.112 | 0.109 | 0.118 |
| **Type B** | | | | |
| Harvard University | United States | 0.254 | 0.221 | 0.170 |
| Stanford University | United States | 0.249 | 0.201 | 0.160 |
| Massachusetts Institute of Technology | United States | 0.235 | 0.196 | 0.155 |
| Yale University | United States | 0.225 | 0.182 | 0.129 |
| Princeton University | United States | 0.211 | 0.177 | 0.125 |
| University of Colorado, Boulder | United States | 0.168 | 0.165 | 0.104 |
| Imperial College London | United Kingdom | 0.161 | 0.135 | 0.120 |
| University of Oxford | United Kingdom | 0.167 | 0.146 | 0.124 |
| University of Cambridge | United Kingdom | 0.170 | 0.145 | 0.129 |
| Lanzhou University | China | 0.098 | 0.074 | 0.052 |
| Fuzhou University | China | 0.127 | 0.121 | 0.107 |
| Katholieke Universiteit Leuven | Belgium | 0.149 | 0.136 | 0.113 |
| Heidelberg University | Germany | 0.147 | 0.127 | 0.116 |
| **Type C** | | | | |
| Nagoya University | Japan | 0.073 | 0.081 | 0.120 |
| Tohoku University | Japan | 0.075 | 0.082 | 0.103 |
| Drexel University | United States | 0.149 | 0.145 | 0.183 |

Data from the CWTS Leiden Ranking 2023, means of four evaluation periods

Next, in each university, comparing the three ratios allowed classifying the universities into three types: A, B, and C, according to the stability or differences between the described ratios. Type A corresponds to stability (deviations up to a maximum of 15% between the highest and lowest values); and Types B and C correspond to a decrease or increase of the ratios, from $P_{top\ 10\%}/P$ to $P_{top\ 1\%}/P_{top\ 10\%}$. Tables 2 and 3 show some examples of universities in the fields of *Physical sciences and engineering* and *Biomedical and health sciences*, respectively, and Table 4 summarizes the number of universities of each type across countries (Tables S1 and S2 include all universities). Most universities are of Type B—53% in both fields—and in a low proportion, they are of Type C: 10% in *Physical sciences and engineering* and 4% in *Biomedical and health sciences*. Consequently, in less than half of the universities the ratios are stable, and therefore they are Type A: 36% in *Physical sciences and engineering* and 43% in *Biomedical and health sciences*.

**Table 3**. Classification of universities attending to the variation of the $P_{top\ 10\%}/P$, $P_{top\ 5\%}/P_{top\ 50\%}$, and $P_{top\ 1\%}/P_{top\ 10\%}$ ratios: stable (Type A), decreasing (Type B), or increasing (Type C). Field of Biomedical and health sciences

| University | Country | $P_{top\ 10\%}/P$ | $P_{top\ 5\%}/P_{top\ 50\%}$ | $P_{top\ 1\%}/P_{top\ 10\%}$ |
|---|---|---|---|---|
| **Type A** | | | | |
| Northeastern University, USA | United States | 0.132 | 0.131 | 0.139 |
| University of Arizona | United States | 0.100 | 0.090 | 0.091 |
| Purdue University, West Lafayette | United States | 0.100 | 0.092 | 0.090 |
| The University of Warwick | United Kingdom | 0.142 | 0.138 | 0.152 |
| University of East Anglia | United Kingdom | 0.146 | 0.132 | 0.129 |
| Peking University | China | 0.086 | 0.079 | 0.080 |
| Xiamen University | China | 0.081 | 0.075 | 0.074 |
| University of Milan | Italy | 0.104 | 0.096 | 0.100 |
| The University of Tokyo | Japan | 0.081 | 0.084 | 0.089 |
| **Type B** | | | | |
| Rockefeller University | United States | 0.326 | 0.282 | 0.182 |
| Massachusetts Institute of Technology | United States | 0.298 | 0.260 | 0.192 |
| Princeton University | United States | 0.267 | 0.236 | 0.173 |
| Yale University | United States | 0.225 | 0.182 | 0.129 |
| University of California, Berkeley | United States | 0.214 | 0.185 | 0.138 |
| California Institute of Technology | United States | 0.213 | 0.185 | 0.150 |
| University of Oxford | United Kingdom | 0.208 | 0.178 | 0.133 |
| University of Cambridge | United Kingdom | 0.208 | 0.174 | 0.130 |
| Tsinghua University | China | 0.137 | 0.126 | 0.115 |
| Soochow University | China | 0.071 | 0.061 | 0.046 |
| University of Toronto | Canada | 0.139 | 0.127 | 0.107 |
| Universität Hamburg | Germany | 0.114 | 0.099 | 0.085 |
| Flinders University | Australia | 0.111 | 0.107 | 0.082 |
| **Type C** | | | | |
| Indiana University, Bloomington | United States | 0.109 | 0.103 | 0.128 |
| Kiel University | Germany | 0.109 | 0.106 | 0.123 |
| Tohoku University | Japan | 0.062 | 0.071 | 0.078 |

Data from the CWTS Leiden Ranking 2023 are means of four evaluation periods

Taking as reference the citation distributions and double rank plots in countries and journals (Rodríguez-Navarro, 2024; also Figures 3 and 4), Type C universities can be explained by a large influence of a lower tail, which contains many more papers that in the reference global distribution. In contrast, Type B universities, in which the $P_{top\ 1\%}/P_{top\ 10\%}$ ratio is lower—in some cases much lower—than the $P_{top\ 10\%}/P$ ratio, show a pattern similar to that of journals with a very high impact factor (Rodríguez-Navarro, 2024). To further investigate Type B universities, I studied the publications of four US universities: Harvard, Stanford, Yale, and the Massachusetts Institute of Technology, in the topic of immunity (as in Figure 1, incorporating these universities into the search). The $P_{top\ 10\%}/P$ and $P_{top\ 1\%}/P_{top\ 10\%}$ ratios corresponding to the retrieved papers were 0.27 and 0.16, respectively, as expected for Type B universities. Figure 6 shows the comparisons of the double rank plot of these publications with the power laws defined by P and $P_{top\ 10\%}$, and $P_{top\ 1\%}$ and $P_{top\ 10\%}$. The results displayed in Figure 6 demonstrate that when calculated from P and $P_{top\ 10\%}$ (using only the less-cited 90% of the papers), the 10% most cited papers have lower ranks than the real ranks. In other words, the predicted research success is higher than it is really. In contrast, when the calculation is preformed with $P_{top\ 1\%}$ and $P_{top\ 10\%}$, the prediction of the extreme of the upper tail is almost correct.

**Table 4**. Number of universities that belong to the three types described in Tables 2 and 3

| | Physical sciences and engineering | | | Biomedical and health sciences | | |
|---|---|---|---|---|---|---|
| **Country** | **Type A** | **Type B** | **Type C** | **Type A** | **Type B** | **Type C** |
| Australia | 4 | 12 | 0 | 7 | 9 | 0 |
| Belgium | 2 | 1 | 0 | 3 | 4 | 0 |
| Brazil | 1 | 0 | 0 | 0 | 0 | 1 |
| Canada | 7 | 0 | 2 | 6 | 5 | 0 |
| China | 23 | 50 | 1 | 9 | 21 | 0 |
| Denmark | 0 | 4 | 0 | 3 | 0 | 0 |
| Finland | 0 | 2 | 0 | 1 | 1 | 0 |
| France | 9 | 2 | 2 | 7 | 6 | 1 |
| Germany | 14 | 9 | 2 | 20 | 3 | 2 |
| Greece | 0 | 1 | 0 | 1 | 0 | 0 |
| India | 0 | 0 | 2 | 0 | 0 | 0 |
| Iran | 1 | 2 | 1 | 0 | 0 | 1 |
| Ireland | 1 | 0 | 0 | 0 | 3 | 0 |
| Israel | 0 | 3 | 1 | 3 | 1 | 0 |
| Italy | 7 | 0 | 0 | 5 | 10 | 0 |
| Japan | 2 | 0 | 6 | 4 | 0 | 1 |
| Malaysia | 0 | 0 | 1 | 0 | 0 | 0 |
| Netherlands | 3 | 4 | 0 | 2 | 7 | 0 |
| New Zealand | 0 | 0 | 1 | 2 | 0 | 0 |
| Norway | 0 | 1 | 0 | 2 | 1 | 0 |
| Poland | 0 | 0 | 1 | 0 | 0 | 1 |
| Portugal | 3 | 0 | 0 | 1 | 0 | 1 |
| Singapore | 0 | 2 | 0 | 1 | 0 | 0 |
| South Korea | 3 | 4 | 4 | 3 | 1 | 0 |
| Spain | 3 | 1 | 2 | 5 | 2 | 0 |
| Sweden | 5 | 1 | 0 | 2 | 3 | 0 |
| Switzerland | 1 | 4 | 0 | 1 | 6 | 0 |
| Taiwan | 0 | 0 | 2 | 2 | 1 | 0 |
| United Kingdom | 7 | 15 | 1 | 8 | 19 | 0 |
| United States | 17 | 48 | 3 | 36 | 61 | 4 |
| **Total** | **113** | **166** | **32** | **134** | **164** | **12** |

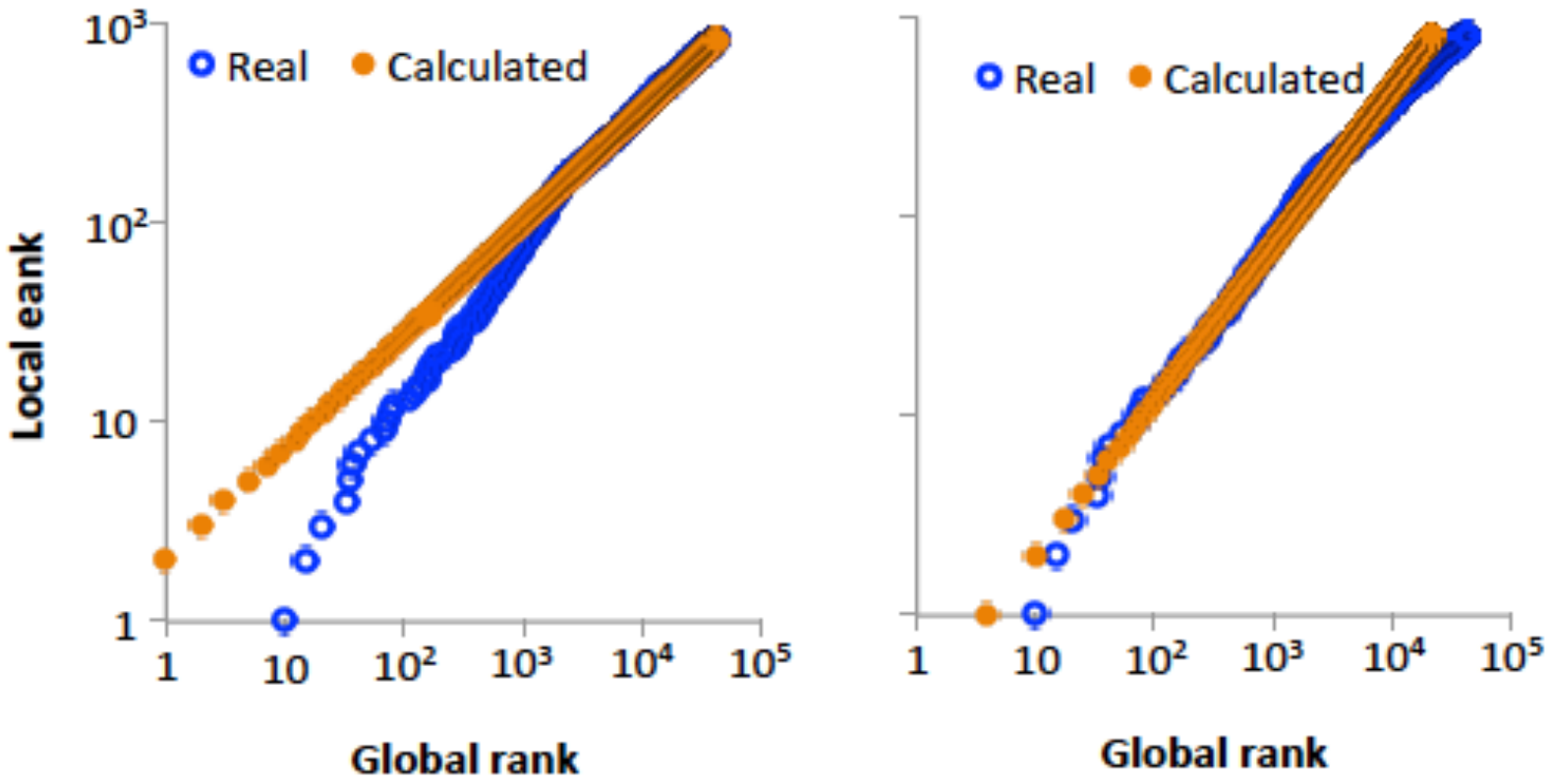

**Figure 6.** Local versus global ranks of US domestic papers published by the universities of Harvard, Stanford, Yale, and Massachusetts Institute of Technology in immunology. Comparison of the ranks of the papers of these universities (local) with those calculated for the power laws defined by P and $P_{top\ 10\%}$ (left), and $P_{top\ 10\%}$ and $P_{top\ 1\%}$ (right)

## 5. Discussion

### 5.1. The essentiality of accurate research assessments

Research not only pushes the boundaries of our knowledge; in current times, it is also a factor that, together with capital and labor, participates in the economic growth of countries (Godin, 2004). Consistent with this fact, developed countries make large investments in research, a significant proportion of which are public investments. Although a certain proportion of the product of this research is technical innovation, which is tangible and measurable in economic terms, a large proportion of its product is knowledge progress, which is intangible and cannot be easily measured. However, a correct research policy must be based on accurate research assessments, and the challenge is to identify the procedure for doing so. This study and a previous one (Rodríguez-Navarro, 2024) show that many current research assessments are not accurate and produce misleading results.

Most research country rankings reported by well-known institutions (European Commission, 2018, 2020, 2022, 2024; National Science Board, 2018, 2020, 2022, 2024) are based on two indicators: one for the size and one for the quality of the system. Currently, the size indicator, number of publications (P), can be considered flaw-free, but the second indicator, based on citations, may be misleading, generating country rankings that do not truly reflect the quality of the research system. Common indicators like $P_{top\ 10\%}/P$ and $P_{top\ 1\%}/P$ define the quality of a research system properly if citations to its publications adhere to the ideal model: a power-law relationship between the citation-based ranks in local and global lists; in other cases, they are misleading. This applies whether rankings aim to provide a general system description, as in most evaluations, or intend to assess the capacity to publish breakthroughs (Gaida et al., 2023).

### 5.2. Deviations in the upper and lower tails of citation distribution

Two types of deviations in the upper and lower tails of citation distributions lead to undervaluing the contribution to pushing the boundaries of knowledge. In the case of the upper tail, a country's capacity is higher than expected from $P_{top\ 10\%}/P$ and $P_{top\ 1\%}/P$ because the most-cited country's papers are more cited than expected from the indicators (e.g., South Korea in solar cells; Figure 1 and 2). This may occur if there is a low proportion of highly competitive researchers with very low influence in bulk production but high influence in the most-cited papers. Currently, there is no information about the frequency of this type of deviation, but in advanced countries, it might not be uncommon.

A more common type of undervaluation is due to an inflated lower tail (Figures 3 and 4). The most misleading effect of this inflated lower tail is on the $P_{top\ 10\%}/P$ and $P_{top\ 1\%}/P$ ratios. The use of these ratios as indicators of research performance implies the assumption that papers with zero or one citation hold no research value. This is the perilous assumption (Rodríguez-Navarro & Brito, 2022) that has caused the inadequate research evaluations of Japan (Rodríguez-Navarro & Brito, 2024a).

This type of effect may be observed in the double-rank analysis (Figure 3, p. 33 in Rodríguez-Navarro, 2024), but it is more easily identified in citation histograms, such as those in Figures 3 and 4, where the proportion of uncited papers or those with only one citation is very different across countries. In this study, inflated lower tails are shown in solar cells and semiconductors (Figure 3) and probably occur in many technological fields.

Although the reason for inflated lower tails has not been definitively demonstrated, in countries with high technology, a reasonable hypothesis is that it is due to the coexistence of two different objectives in the research system: pushing the boundaries of knowledge and incremental innovations. The papers with the latter purpose are poorly cited, and many are uncited even in wide citation windows. This infrequent citation contrasts sharply with research focused on pushing the boundaries of knowledge, whose publications rarely remain uncited (Rodríguez-Navarro & Brito, 2022).

**5.3. Deviations from the global distribution**

In addition to the deviations in the upper and lower tails just described, two other types of deviations lead to overvaluing the contribution to pushing the boundaries of knowledge. In these cases, the $P_{top\ 10\%}/P$, $P_{top\ 5\%}/P_{top\ 50\%}$, and $P_{top\ 1\%}/P_{top\ 10\%}$ ratios show decreasing values.

One type of deviation affects the whole distribution, which is displaced to the right in the histograms with logarithmic binning because the citation lognormal distribution of the country or institution has a much higher $\mu$ parameter than the global distribution. In isolation, his type of deviation can be observed with synthetic series but may be infrequent in real cases.

In the other type, the deviation from the lognormal distribution occurs because the lower tail of the citation distribution is decreased with reference to the upper tail. In other words, “normal” research is partially excluded in these cases. These notable deviations from the global distribution occur in journals with a very high impact factor (Rodríguez-Navarro, 2024), and it is reproduced in Type B universities.

Many well-known, research-intensive universities (e.g., Harvard, Stanford, Yale, Cambridge, or Oxford; Tables 2, 3, S1, and S2) fall into the Type B category. A reasonable explanation for this finding is that the number of lowly funded, less competitive researchers in these universities is much lower than in global research. Possibly, across the world, the number of Type B universities may not be very high. The high proportion of Type B universities in this study is probably the result of the selection of universities with $P_{top\ 1\%}$ higher than nine in the four evaluation periods. This implies a stringent selection of universities and, consequently, an increase in the proportion of Type B universities in the selection. In any case, it seems that the proportion of these universities is notably different across countries (Table 4).

In these universities, the $P_{top\ 10\%}/P$ and $P_{top\ 1\%}/P$ ratios overvalue the capacity to push the boundaries of knowledge. This capacity is correctly valued by $P_{top\ 10\%}/P_{top\ 1\%}$ (Figure 6). For example, in Physical Sciences and Engineering (Table 2), in the USA, the $P_{top\ 1\%}/P_{top\ 10\%}$ ratio of Florida State University is similar to those of Yale University and Princeton University, and the same occurs in the UK with the universities of Nottingham and Exeter compared to those of Oxford or Cambridge. In the absence of deviations in the extreme upper tail (South Korea in Figure 2), the similar $P_{top\ 1\%}/P_{top\ 10\%}$ ratios of these universities indicate similar contributions to pushing the boundaries of knowledge. In contrast, if the evaluation is based on the $P_{top\ 10\%}/P$ ratio, we would reach the conclusion that Florida State University in the USA and the universities of Nottingham and Exeter in the UK are second-tier institutions compared to Yale University, Princeton University, and Oxford and Cambridge Universities. Focusing on the contribution to pushing the boundaries of knowledge, this would be an inaccurate and unfair conclusion, which might extend to many apparent second-tier institutions and even countries.

### 5.4. The Need for a Revised Evaluation Paradigm

Building upon a previous report (Rodríguez-Navarro, 2024), the present study highlights the complexity of research evaluations across various countries and institutions. An effective and informed research policy requires the consideration of four key indicators of research (Rodríguez-Navarro, 2024), in conjunction with an analysis of both disruptive and incremental innovations. However, it seems reasonably that large-scale assessments of national performance in science, engineering, and innovation—such as those conducted by the European Commission (e.g., European Commission, 2024), the National Science Board (e.g., National-Science-Board, 2024), Hcéres (Hcéres, 2019), and the National Institute of Science and Technology in Tokyo (National Institute of Science and Technology, 2022)—adopt a single research efficiency indicator that reflects the capacity of the research system to advance the frontiers of knowledge.

In this context, it is crucial to move away from the widespread use of the $P_{top\ 10\%}/P$ and $P_{top\ 1\%}/P$ ratios for size-independent research assessments. First, it is important to note that not all research is aimed at advancing the frontier of knowledge, making the use of P as a normalizing factor inappropriate. The top 10% and top 1% most cited papers belong exclusively to this type of research, while P includes all types of research, meaning that the ratios are influenced by differences in research focus and citation patterns of other types of research. Additionally, in some leading research institutions, papers with fewer citations are disproportionately underrepresented compared to those from other countries and institutions. As a result, while the $P_{top\ 10\%}$ and $P_{top\ 1\%}$ have similar meaning in global and local citation distributions, P is sensitive to variations in the publication of lowly cited papers.

Even if the proportion of misleading evaluations based on $P_{top\ 10\%}/P$ and $P_{top\ 1\%}/P$ is low, their application—without transparency regarding deviations from the ideal model—generates uncertainty in all instances. The persistent classification of Japan as an underdeveloped country in terms of research, despite its substantial achievements, serves as a clear cautionary example of the misapplication of these metrics. A more appropriate alternative is the $P_{top\ 1\%}/P_{top\ 10\%}$ ratio, which can be derived from the same datasets currently used, thus ensuring consistency while addressing the limitations of existing indicators. If $P_{top\ 1\%}$ is too low, $P_{top\ 5\%}/P_{top\ 25\%}$ or in extreme cases $P_{top\ 10\%}/P_{top\ 50\%}$ may be reliable alternatives.

In size-dependent assessments of breakthroughs in specific research topics, the relatively low number of publications per topic means that the top 0.01% cited papers cannot be used due to statistical constraints, and the $P_{top\ 10\%}$ is typically employed (Gaida et al., 2023; Leung et al., 2024). In such cases, calculating the probability of publishing in the top 0.01% based on the $P_{top\ 5\%}/P_{top\ 25\%}$ ratio is not advisable, given the deviations observed at the extreme upper tail (Figures 1 and 2). A more tailored indicator is necessary to accurately capture the performance of research at the extreme of the upper tail of citation distribution (Rodríguez-Navarro & Brito, 2024b).

## 6. Conclusions and Implications

A significant number of current country and institution rankings, as well as evaluations conducted by well-known organizations, rely on research efficiency indicators, particularly the $P_{top\ 10\%}/P$ or $P_{top\ 1\%}/P$ ratios. Both prior and current studies demonstrate that these ratios often fail to provide a reliable measure of research efficiency in various contexts. Therefore, the continued use of such indicators is not recommended. The reliance on inappropriate evaluation metrics in research policy could lead to misguided decision-making, ultimately impairing the effectiveness of research strategies and their outcomes.

## CONFLICT OF INTEREST

The author declares that there is no conflict of interest.

## REFERENCES

Abramo, G., Cicero, T., & D'Angelo, C. A. (2013). National peer-review research assessment execises for the hard sciences can be a complete waste of money: the Italian case. *Scientometrics*, *95*, 311-324.

Aksnes, D. W., Langfeldt, L., & Wouters, P. (2019). Citations, citation indicators, and research quality: An overview of basic concepts and theories. *SAGE Open*, *January-March: 1-17*.

Bihari, A., Tripathi, S., & Deepak, A. (2021). A review on *h*-index and its alternative indices. *Journal of Information Science*, 1-42.

Bornmann, L., & Daniel, H.-D. (2007). What do we know about the *h* index. *Journal of the American Society for information Science and Technology*, *58*, 1381-1385.

Bornmann, L., Leydesdorff, L., & Mutz, R. (2013). The use of percentile rank classes in the analysis of bibliometric data: opportunities and limits. *Journal of Informetrics*, *7*, 158-165.

Bornmann, L., Ye, A., & Ye, F. (2018). Identifying landmark publications in the long run using field-normalized citation data. *Journal of Documentation*, *74*, 278-288.

Brito, R., & Rodríguez-Navarro, A. (2021). The incosistency of *h*-index: A mathematical analysis. *Journal of Informetrics*, *15*, Article 101106.

Conover, W. J., & Iman, R. (1981). Rank transformations as a bridge between parametric and nonparametric statitics. *The American Statistician*, *35*, 124-129.

Docampo, D., & Besoule, J.-J. (2019). A new appraoch to the analysis and evaluation of the research output of countries and institutions. *Scientometrics*, *119*, 1207-1225.

Drucker, P. F. (1954). *The Practice of Management*. Harper and Row Publishers, New Yrok.

European Commission. (2018). *Science, Research and Innovation Performance of the EU 2018. Strengthening the foundations for Europe's future*. Publications Office of the European Union.

European Commission. (2020). *Science, Research and Innovation Performance of the EU 2020. A fair, green and digital Europe*. Publication Office of the European Union.

European Commission. (2022). *Science, Reserach and Innovation Performance of the EU. Building a sustainable future in uncertain times*. European Commision.

European Commission. (2024). *Science, Reserach and Innovation Performance of the EU. A competitive Europe for a sustainable future*. European Commission.

Gaida, J., Wong-Leung, J., Robin, S., & Cave, D. (2023). *ASPI's Critical Technology Tracker. The global race for future power*. Policy Brief Report No.69/2023

Garfield, E. (1955). Citation indexes for science. *Science*, *122*, 108-111.

Garfield, E. (1973). Citation frequency as a measure of research activity and performance. *Essays of an Information Scientist*, *1*, 406-408.

Garfield, E. (1999). Journal Impact Factor: a brief review. *Canadian Medical Association journal*, *161*, 979-980.

Garfield, E., & Welljams-Dorof, A. (1992). Citation data: their use as quantitative indicators for science and technology evaluations and policy-making. *Science and Public Policy*, *19*, 321-327.

Godin, B. (2003). The emergence of S&T indicators: why did governments supplement statistics with indicators? *Research Policy*, *32*, 679-691.

Godin, B. (2004). The new economy: what the concept owes to the OECD. *Research Policy*, *33*, 679-690.

Godin, B. (2006). On the origins of bibliometrics. *Scientometrics*, *68*, 109-133.

Golosovsky, M. (2021). Universality of citation distributions: A new understanding. *Quantitative Science Studies*, *2*, 527-543.

Hcéres. (2019). *OST, Dynamics of scietific production in the world, in Europe and France, 2000-2016*. Paris.

Hirsch, J. E. (2005). An index to quantify an individual's scientific reserach output. *Proceedins of the National Academy of Sciences USA*, *102*, 16569-16572.

Hu, X., & Rousseau, R. (2019). Do citation chimeras exist? The case of under-cited influential articles suffering delayed recognition. *Journal of the Association for Information Science and Technology*, *70*, 499-508.

Kuhn, T. (1970). *The structure of scientific revolutions*. University of Chicago Press.

Leung, J. W., Robin, S., & Cave, D. (2024). *Techology Tracker: The rewards of long-term reserach investments*.

Mcalister, P. R., Narin, F., & Corrigan, J. G. (1983). Programmatic evaluation and comparison based on standardized citatio scores. *IEEE Transactions on Engineering Managament*, *EM-30*, 205-2011.

Merton, R. K. (1979). *Foreword. In E. Garfield (Ed), Citation Indexing - Its Theory and Application in Science, Technology, and Humanities*. John Willey & Sons.

Min, C., Bu, Y., Wu, D., Ding, Y., & Zhang, Y. (2021). Identifying citation patterns of scientific breakthroughs: A perspective o dynamic citation process. *Information Processing & Management*, *58*, 102428.

Mingers, J., & Leydesdorff, L. (2015). A review of theory and practice in scientometrics. *European Journal of Operational Research*, *246*, 1-19.

Narin, F. (1976). *Evaluative bibliometrics: The Use of Publication and Citation Analysis in the Evaluation of Scientific Activity*. Computer Horizon Inc.

National Institute of Science and Technology, T. (2022). *"Japanese Science and Technology Indicators 2022", NISTEP RESEARCH MATERIAL No. 318*.

National Science Board. (2018). *Science and Engineering Indicators 2018*. National Science Foundation.

National Science Board. (2020). *Science and Engineerin Indicators 2020: The State of U.S. Science and Engineering*. NSB-2020-1. Alexandria, VA.

National Science Board. (2022). *Science and Enginering Indicators 2022: The State of U.S. Science and Engineering. NSB-2022-1*.

National Science Board. (2024). *Science and Engineering Indicators 2024: The State of U.S. Science and Engineering. NSB-2024-3*.

Olensky, M., Schmidt, M., & van Eck, N. J. (2015). Evaluation of the citation matching algorithms of CWTS and iFQ in comparison to the Web of Science. *Journal of the Association for Information Science and Technology*, *67*, 2550-2564.

Pendlebury, D. A. (2020). When the data don't mean what they say: Japan's comparative underperformance in citation impact. In C. Daraio & W. Glanzel (Eds.), *Evaluative Informetrics: The Art of Metrics-based Research Assessment*. Spriger.

Perianes-Rodriguez, A., & Ruiz-Castillo, J. (2016). University citation distribution. *Journal of the Association for Information Science and Technology*, *67*, 2790-2804.

Poege, F., Harhoff, D., Gaessler, F., & Baruffaldi, S. (2019). Science quality and the value of inventions. *Science Advances*, *5*, eaay7323.

Prathap, G. (2010). An iCE map approach to evaluate performance and efficiency ofscientific production of countries. *Scientometrics*, *85*, 185-191.

Price, D. J. d. S. (1965). Networks of scientific papers. *Science*, *149*, 510-515.

Radicchi, F., Fortunato, S., & Castellano, C. (2008). Universality of citation distributions: toward an objective measure of scientific impact. *Proceedins of the National Academy of Sciences USA*, *105*, 17268-17272.

Rodríguez-Navarro, A. (2024). Citation distributions and reserach evaluations: The impossibility of formulating a universal indicator. *Journal of Data and Information Science*, *9*, 24-48.

Rodríguez-Navarro, A., & Brito, R. (2018a). Double rank analysis for research assessment. *Journal of Informetrics*, *12*, 31-41.

Rodríguez-Navarro, A., & Brito, R. (2018b). Technological research in the EU is less efficient than the world average. EU research policy risks Europeans' future. *Journal of Informetrics*, *12*, 718-731.

Rodríguez-Navarro, A., & Brito, R. (2019). Probability and expected frequency of breakthroughs – basis and use of a robust method of research assessment. *Scientometrics*, *119*, 213-235.

Rodríguez-Navarro, A., & Brito, R. (2020). Like-for-like bibliometric substitutes for peer review: advantages and limits of indicators calculated from the $e_p$ index. *Research Evaluation*, *29*, 215-230.

Rodríguez-Navarro, A., & Brito, R. (2021b). Total number of papers and in a single percentile fully describes research impact – Revisiting concepts and applications. *Quantitative Science Studies*, *2*, 544-559 .

Rodríguez-Navarro, A., & Brito, R. (2022). The link between countries' economic and scientific wealth has a complex dependence on technological activity and research policy. *Scientometrics*, *127*, 2871-2896.

Rodríguez-Navarro, A., & Brito, R. (2024a). The extreme upper tail of Japan's citation distribution reveals its research success. *Quality & Quantity*, *58*, 3831-3844.

Rodríguez-Navarro, A., & Brito, R. (2024b). Rank analysis of most cited publications, a new approach for research assessments. *Journal of Informetrics*, *18*, 101503.

Schlagberger, E. M., Bornmann, L., & Bauer, J. (2016). At what institutions did Nobel lauretae do their prize-winning work? An analysis of bibliographical information on Nobel laureates from 1994 to 2014. *Scientometrics*, *109*, 723-767.

Schneider, J. W., & Costas, R. (2017). Identifying potential "breakthrough" publications using refined citation analyses: Three related explorative approaches. *Journal of the Association for Information Science and Technology*, *68*, 709-723.

Schubert, A., & Braun, T. (1996). Cross-field normalization of scientometric indicators. *Scientometrics*, *36*, 311-324.

Shahmandi, M., Wilson, P., & Thelwall, M. (2020). A new algorithm for zero-modified models applied to citation counts. *Scientometrics*, *125*, 993-1010.

Thelwall, M. (2016). Are there too many articles? Zero inflated variants of the discretised lognormal and hooked power law. *Journal of Informetrics*, *10*, 622-633.

Thelwall, M., Kousha, K., Stuart, E., Makita, M., Abdoli, M., Wilson, P., & Levitt, J. (2023). In which fields are citations indicators of reserach quality? *Journal of the Association for Information Science and Technology*, *74*, 941-953.

Traag, V. A., & Waltman, L. (2019). Systematic analysis of agreement between metrics and peer review in the UK REF. *Palgrave Communications*, *5*, 29. https://doi.org/ https://doi.org/10.1057/s41599-019-0233-x

Viiu, G.-A. (2018). The lognormal distribution explains the remarkable pattern documented by characteristic scores and scales in scientometrics. *Journal of Informetrics*, *12*, 401-415.

Waltman, L. (2016). A review of the literature on citation impact indicators. *Journal of Informetrics*, *10*, 365-391.

Waltman, L., Calero-Medina, C., Kosten, J., Noyons, E. C. M., Tijssen, R. J. W., van Eck, N. J., van Leeuwen, T. N., van Raan, A. F. J., Visser, M. S., & Wouters, P. (2012). The Leiden ranking 2011/2012: Data collection, indicators, and interpretation. *Journal of the American Society for information Science and Technology*, *63*, 2419-2432.

Waltman, L., & Schreiber, M. (2013). On the calculation of percentile-based bibliometric indicators. *Journal of the American Society for information Science and Technology*, *64*, 372-379.

Waltman, L., & van Eck, N. J. (2012). The inconsistency of the h-index. *Journal of the American Society for information Science and Technology*, *63*, 406-415.

Waltman, L., van Eck, N. J., van Leeuwen, T. N., Visser, M. S., & van Raan, A. F. J. (2011). Towards a new crown indicator: Some theoretical considerations. *Journal of Informetrics*, *5*, 37-47.

Waltman, L., van Eck, N. J., & van Raan, A. F. J. (2012). Universality of citation distributions revisited. *Journal of the American Society for information Science and Technology*, *63*, 72-77.

Wildgaard, L., Schneider, J. W., & Larsen, B. (2014). A rewiew of the characteristics of 108 author-level bibliometric indicators. *Scientometrics*, *101*, 125-158.

Wuestman, M., Hoekman, J., & Frenken, K. (2020). A topology of scientific breakthroughs. *Quantitative Science Studies*, *1*, 1203-1222.